\documentclass[useAMS,usenatbib,usegraphicx]{mn2e}
\usepackage{epsfig}

\def\kms{km~s$^{-1}$}

\begin{document}

\title[Stellar velocity dispersion in NLS1 galaxies]
{Stellar velocity dispersion in NLS1 galaxies
\thanks{Based on observations collected at Asiago observatory.}
}

\author[Botte et al.]{V. Botte$^{1,2}$, S. Ciroi$^{1}$, F. Di Mille$^{1}$, 
P. Rafanelli$^{1}$ and A. Romano$^{1}$\\
$^{1}$ Department of Astronomy, University of Padova, vicolo dell'Osservatorio 2, I-35122 Padova, Italy\\
$^{2}$ Guest investigator of the UK Astronomy Data Centre
}

\date{Accepted 2004 October 15}
%      Received 2004 July 1;
%      in original form 1988 October 11}
%\pagerange{\pageref{firstpage}--\pageref{lastpage}}
%\pubyear{1994}

\maketitle
\label{firstpage}

\begin{abstract}
Several authors recently explored the Black-Hole mass ($M_{BH}$) vs. stellar 
velocity dispersion ($\sigma_*$) relationship for Narrow Line Seyfert 1 
galaxies (NLS1s). Their results are
more or less in agreement and seem to indicate that NLS1s fill the region below
the fit obtained by \citet{tret02}, showing a range of $\sigma_*$ similar to
that of Seyfert 1 galaxies, and a lower $M_{BH}$.
Until now the [O\,{\sc iii}] width has been used in place of the stellar 
velocity dispersion, 
but some indications begin to arise against the effectiveness of the
gaseous kinematics in representing the bulge potential at least in NLS1s.
\citet{bz04b} stressed the urgency to produce true $\sigma_*$ measurements.
Here we present new stellar velocity dispersions obtained through direct
measurements of the Ca\,{\sc ii} absorption triplet ($\sim 8550$ \AA) in the 
nuclei of 8 NLS1 galaxies. The resulting $\sigma_*$ values and a comparison 
with $\sigma_{\rm [O~III]}$ confirm our suspects (see \citealt{bo04}) that
[O\,{\sc iii}] typically overestimates the stellar velocity dispersion and 
demonstrate that NLS1s follow the $M_{BH}-\sigma_*$ relation as Seyfert 1, 
quasars and non-active galaxies.
\end{abstract}


\section{Introduction}
Recently great interest was addressed to the study of the correlation between 
the mass of nuclear supermassive Black-Holes ($M_{BH}$) and the stellar velocity 
dispersion ($\sigma_*$) of their hosting bulges.
This relation has been obtained for nearby non-active galaxies by several 
authors \citep{gebet00a, fermer00, tret02, mahu03}, supporting the theory that 
the growth of the BH is bound to the galaxy formation. According to \citet{dima03} 
this correlation, not set in primordial structures but fully established 
only at low redshift, can be understood if there is a simple linear relation 
between the total gas mass in galaxies and their $M_{BH}$.\\
\citet{gebet00b} included in their sample seven Active Galactic Nuclei (AGNs)
and found that also these objects followed a similar $M_{BH}-\sigma_*$ 
correlation of non-active galaxies. Other authors have then confirmed the 
validity of this result for samples of Seyfert 1 galaxies (S1s) and quasars 
\citep{ne00, fe01, wan02, bo03, sh03}. 
On the contrary, no consensus has been reached 
until now about the case of the Narrow-Line Seyfert 1 galaxies (NLS1s).
These AGNs have spectroscopic properties slightly different from 
those of the classical S1s, as for example, narrower permitted 
emission lines in the optical wavelength domain and steeper power-law X-rays 
continua, which suggest that NLS1s are hosting BHs with smaller masses accreting 
at high rates, close to the Eddington limit. This hypotesis would imply that 
NLS1s have also less massive bulges than classical S1s, of course assuming that 
their $M_{BH}$ are correlated to the physical properties of their hosting 
bulges. 
Indeed, \citet{walu01}, \citet{wan02} and \citet{wan04}  
found that there is no clear 
difference in the $M_{BH}-\sigma_*$ relation between narrow-line, broad-line 
AGNs and non-active nearby galaxies, while opposite results were obtained by 
\citet{matet01}, \citet{bz04a}, \citet{gm04}, and \citet{bo04}. 
The main problem is that $\sigma_*$ in AGNs is available for few S1s 
and only 3 NLS1s: Mrk~110 ($90\pm 7$ \kms) by \citet{fe01}, NGC~4051 
($88\pm 13$ \kms) by \citet{nel95}, and Mrk~766 ($106 \pm 40$ \kms) by \citet{JB00}.  
In place of $\sigma_*$, the [O\,{\sc iii}] emission linewidth was extensively 
used as a representation of the bulge velocity dispersion.

Here we present new direct determinations of  $\sigma_*$ in the nuclei 
of 8 NLS1 galaxies, obtained using the Ca\,{\sc ii} absorption triplet 
(8498, 8542, 8662 \AA). 
The targets were extracted from \citet{bo04}, except for Mrk~766.
In Section 2 we present the spectroscopic observations and the data reduction; 
in Section 3 we give a detailed description of the methods we used to measure
stellar and gaseous kinematics, and we present the results.
Finally, in Section 4 we discuss the implications of our results on the validity
of the FWHM([O\,{\sc iii}])/2.35=$\sigma_*$ and $M_{BH}-\sigma_*$ relations for 
NLS1 galaxies.

\begin{figure*}
\includegraphics[angle=0,width=80mm]{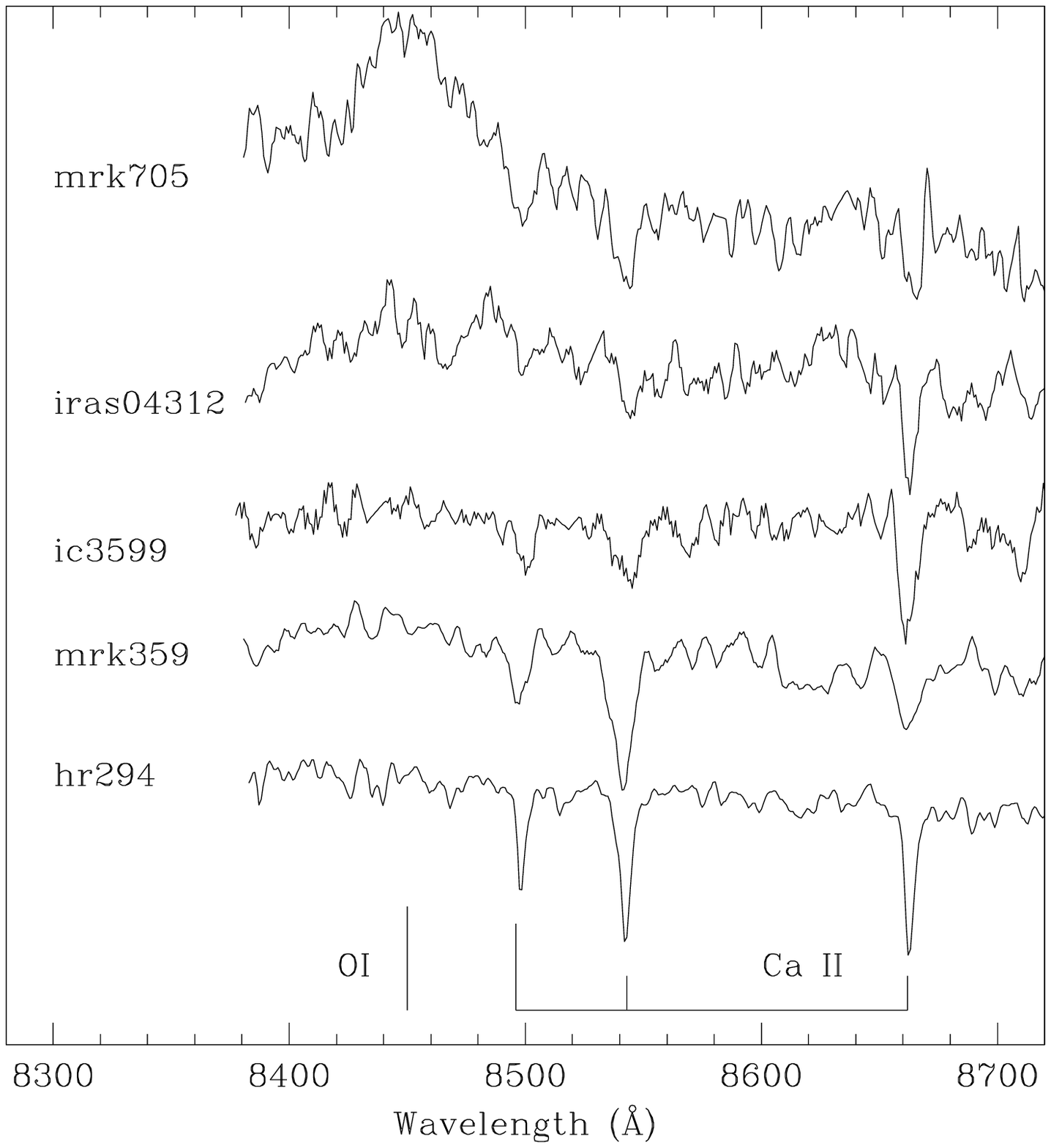}
\includegraphics[angle=0,width=80mm]{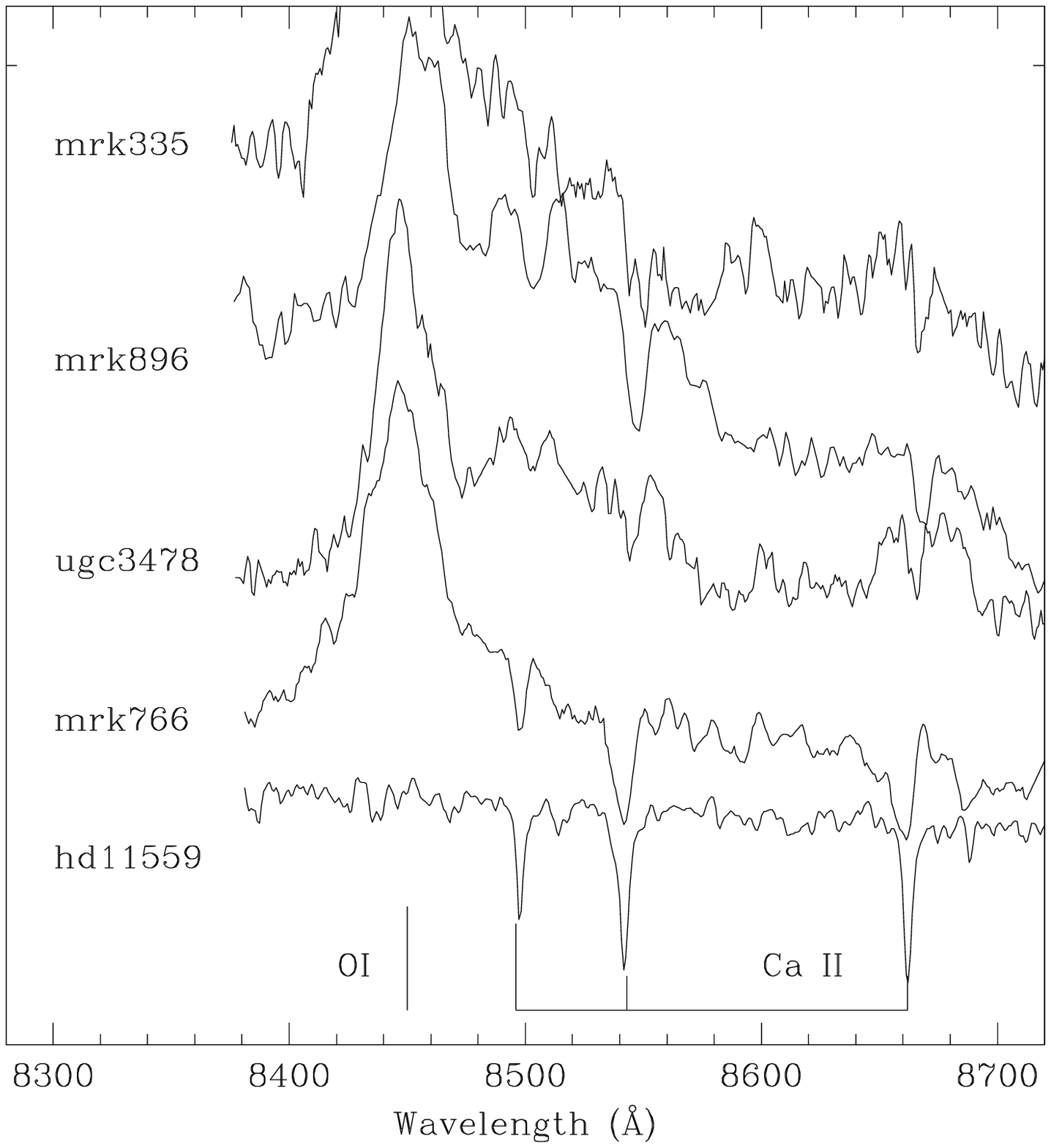}
 \caption{The rest-frame spectra of the 8 NLS1 galaxies showing the Ca\,{\sc ii} 
 triplet absorption lines. The template stars (HR 294 and HD 11559) are also 
 plotted for comparison.}
\label{fig1}
\end{figure*}

\section{OBSERVATIONS AND DATA REDUCTION}

We have observed the nuclei of 7 NLS1s with the Asiago Faint Object 
Spectrograph Camera (AFOSC) mounted at the 1.82 m telescope of the Padova 
Astronomical Observatory (Asiago, Italy). A Volume Phase Holographic grism 
was used in combination with a 1.69 arcsec-slit to cover 
the region $\sim$ 8200--9200 \AA\, which includes the Ca\,{\sc ii} triplet 
absorption lines, 
with a dispersion of $\sim 0.88$ \AA/px and a relatively high spectral 
resolution ($\sim 50$ \kms). 
Mrk~766 was also observed in the optical wavelength range (4300--8000 \AA) at 
medium resolution, in order to derive the $M_{BH}$ and [O\,{\sc iii}] width 
information not availble in \citet{bo04} and useful for the following 
analysis.
In addition to galaxy observations, the spectra of two template stars, HR~294 
(G9 III) and HD~11559 (K0 III),  were obtained for kinematic analysis with the 
same configuration. Two spectral types were chosen to minimize the template
mismatch effects.

Mrk~896 was extracted from the Isaac Newton Group (ING) Archive.
A spectrum covering a larger range ($\sim$ 7700--9400 \AA) was obtained in August
1994 at the William Herschel Telescope (Canary Islands, Spain) with ISIS Double
Beam Spectrograph.
The R316R grating used in combination with a 1.47 arcsec-slit yielded a 
dispersion of $\sim$ 1.35 \AA/px and an instrumental resolution of 
$\sim$ 65 \kms.
A standard star (BD+17 4708) was observed during the same run for
spectrophotometric calibration.  

The spectra are shown in Fig.~\ref{fig1}. Details about the observations are 
given in Table \ref{tab1}.\\
All data were reduced with the same procedure.
The usual reduction steps, that is bias and flat field corrections, cosmic rays 
removal, wavelength calibration by means of comparison lamps 
and sky-background subtraction, were carried out with 
IRAF packages\footnote{IRAF is written and supported by NOAO (Tucson, Arizona), 
which is operated by AURA, Inc. under cooperative agreement with the National 
Science Foundation}.
The optical spectrum of Mrk~766 was also flux calibrated through the observation
of two spectrophotometric standard stars, FEIGE56 and G191-B2B.

For each object we extracted the nuclear spectrum summing a number of pixels
along the slit to increase the signal-to-noise ratio (S/N). 
An aperture of $\sim 3.3$ arcsec (7 px) was chosen for the Asiago data on the 
basis of the seeing, which ranged between 2 and 3 arcsec during the 
observations.
This aperture corresponds to $\sim 0.6-2$ kpc for galaxies in the range 
$0.01<z<0.03$.
For the ING spectrum we used an aperture of 1.7 arcsec (5 px), corresponding to 
$\sim 0.9$ kpc, following similar arguments.

\begin{table}
\caption{Observation Log. \label{tab1}}
\begin{tabular}{l c c c c c }
\hline 
Galaxy & z & $\Delta\lambda$ &  Date & T$_{exp}$  \\
&  & (\AA) & & (sec)  \\
\hline
Mrk 359	 & 0.017  & 8200--9200 & 2003-09-08 & 7200  \\
Mrk 335  & 0.026  & 8200--9200 & 2003-12-07 & 9000 \\ 
UGC 3478 & 0.013  & 8200--9200 & 2003-12-08 & 7200  \\
Mrk 705	 & 0.029  & 8200--9200 & 2003-12-08 & 7200  \\
IR 04312 & 0.020  & 8200--9200 & 2004-01-26 & 7200  \\
IC 3599  & 0.022  & 8200--9200 & 2004-01-26 & 7200  \\
Mrk 766  & 0.013  & 4300--6500 & 2004-01-29 & 2400 \\
	 &	  & 6200--8000 & 2004-01-30 & 1800 \\
	 & 	  & 8200--9200 & 2004-01-30 & 7200  \\
	 &        &            &            &       \\
Mrk 896  & 0.026  & 7700--9500 & 1994-08-16 & 1800 \\
\hline
\end{tabular}
\end{table}

\section{DATA ANALYSIS}

\subsection{Stellar velocity dispersion}
To measure $\sigma_*$ we have used the Fourier cross-correlation method as 
implemented in the IRAF task \textit{FXCOR} which is based on the method of 
\citet{toda79}. The spectrum of the galaxy G and the stellar template T 
are resampled into N bins, where each bin number $n$ is proportional to 
$\ln~\lambda$, and cross-correlated in Fourier space:

\begin{equation}
C(n) = G(n) \times T(n) = \int_{-\infty}^\infty T(k)G(k+n)dk
\end{equation}
where $\times$ means cross correlation.

This method assumes that the G(n) is a convolution of T(n),  
representing the true stellar population in the galaxy, with a broadening 
function b(n), offset from zero velocity by an amount $\delta$:

\begin{equation}
g(n)\propto b(n-\delta)\ast t(n) = \int_{-\infty}^{\infty} b(x-\delta)t(n-x)dx
\end{equation}
where $\ast$ denotes convolution. 

The maximum peak of the cross-correlation function (CCF) is fitted by a smooth 
symmetric function (in general a gaussian function), whose location and width 
are related to the galaxy redshift and velocity dispersion of the stars 
in the galaxy.
In order to recover $\sigma_*$ from the FWHM of the CCF peak, we adopted
the suggestions given by \citet{nel95}. We convolved the template spectrum with 
gaussian functions of known increasing dispersions (from 10 to 300 \kms) and we 
measured the corresponding increasing widths of the CCF peak. 
Then, we fitted the empirical FWHM-$\sigma_*$ relationship with a polynomial 
function.

Before to run \textit{FXCOR}, the spectra of the NLS1s were moved to the
rest-frame and 
their continuum carefully removed. This last step proved to be rather important, 
since the cross-correlation method is sensible to low-frequency fluctuations 
and our targets are often characterized by emission lines and/or bumps.
In particular, 5 out of the 8 observed targets show the O\,{\sc i}
$\lambda$8446 emission
line, and in at least 3 galaxies (Mrk 335, Mrk 896, UGC 3478) the Ca\,{\sc ii}
absorptions are overlapped to broader Ca\,{\sc ii} emissions. 
Then, \textit{FXCOR} was run using two template stars for the Asiago data, and 
the only one available for the ING archival data. 
The resulting CCF peaks had values of the Tonry-Davis $R$
parameter \citep{toda79}, a sort of S/N indicator, ranging from 8 to 20.  
No filtering of high-frequency noise was 
applied, since we verified that it was not necessary and introduced variations 
of the results depending on the arbitrary choice of the filter shape.

To estimate the uncertainty of the $\sigma_*$ measurements, we 
followed \citet{Oh02}. Firstly, two spectral regions were independently 
correlated for each template spectrum: 8300--8800 \AA~ and 8450--8750 \AA.
This allowed to have four values of $\sigma_*$ for each galaxy. We adopted  
the mean of these measurements and calculated the RMS ($\delta\sigma_1$).
$\delta\sigma_1$ was not estimated for Mrk 896, since only one template star was
available.
Then, we made 20 spectra for each object by adding artificial noise with the
IRAF task \textit{MKNOISE} to increase the RMS by $\sim20$ per cent. 
\textit{FXCOR} was run again 20 times with these
new spectra and the scatter of the peak width measurements was adopted as error
of the CCF peak fitting ($\delta\sigma_2$).
Finally, the uncertainty of the instrumental resolution ($\delta\tau\sim$2--4
\kms) was considered as additional source of error by deriving the Eq. (18) by
\citet{toda79}: $\delta\sigma_3=2\tau\delta\tau/\sigma_*$.
Assuming these errors are independent one from each other, the total
$\Delta\sigma_*$ was obtained through the following relation:  
$\Delta\sigma_*=\sqrt{(\delta\sigma_1)^2+(\delta\sigma_2)^2+(\delta\sigma_3)^2}$.
The stellar velocity dispersions are listed in Table \ref{tab2}.

\subsection{Gas velocity dispersion}
In addition to measurements of stellar kinematics from absorption 
lines, we took from our previous work \citep{bo04} the optical spectra of the 
same objects, containing the Narrow Line Region (NLR) emission 
lines which allow to measure the gas kinematics. In particular, the width of 
[O\,{\sc iii}] $\lambda$5007 line was measured and then corrected for 
instrumental line broadening to determine the velocity dispersion of the gas 
in the NLR. 
Since the [O\,{\sc iii}] emission line is bright in these spectra, its profile 
is very little affected by noise. 
Therefore, we fitted [O\,{\sc iii}] with gaussian functions 
only changing for 5 times the continuum level, chosen by visual inspection.
The average dispersion $\sigma_{\rm [O~III]}$ and its relative scatter were 
calculated for each galaxy.
The gas velocity dispersions are given in Table \ref{tab2}.

\begin{table}
%\centering
\begin{minipage}{10cm}
\caption{Stellar and gaseous velocity dispersions, and BH masses.} 
\label{tab2}
\begin{tabular}{lccc}
\hline 
Galaxy & $\sigma_*$ & $\rm \sigma_{[O~III]}$ & $M_{BH}$ \\
 & (\kms) &  (\kms) & ($\times 10^6~ M_{\sun}$) \\
\hline
Mrk 335	 &  64 $\pm$ 17	 & 140.8 $\pm$ 3.4 & 8.61 \\
Mrk 359	 & 112 $\pm$ 12  &  67.0 $\pm$ 0.4 & 0.69 \\
UGC 3478 &  89 $\pm$ 18  & 109.8 $\pm$ 2.6 & 0.81  \\
Mrk 705	 &  82 $\pm$ 25  & 185.9 $\pm$ 2.1 & 46.13 \\
IR0 4312 &  64 $\pm$ 20  & 136.8 $\pm$ 7.8 & 0.83 \\
IC 3599  &  85 $\pm$ 17  & 106.9 $\pm$ 4.4 & 0.13 \\
Mrk 766  &  81 $\pm$ 17  & 131.2 $\pm$ 0.7 & 0.63 \\
Mrk 896	 &  87 $\pm$ 11	 & 129.6 $\pm$ 0.6 & 4.43 \\
& & & \\
Mrk 110	 &  90 $\pm$ 7$^a$  & 123.3$^b$ & 6.65$^c$ \\
NGC 4051 &  88 $\pm$ 13$^d$  & 81$^b$    & 1.35$^c$ \\
\hline
\end{tabular}
\end{minipage}

\begin{tabular}{l}
$^a$ Average value from \citet{fe01} \\
$^b$ From \citet{whi92}\\
$^c$ Average value from \citet{ka00}\\
$^d$ From \citet{nel95}\\
\end{tabular}
\end{table}

\subsection{Results}
 
\begin{figure}
\begin{center}
\psfig{file=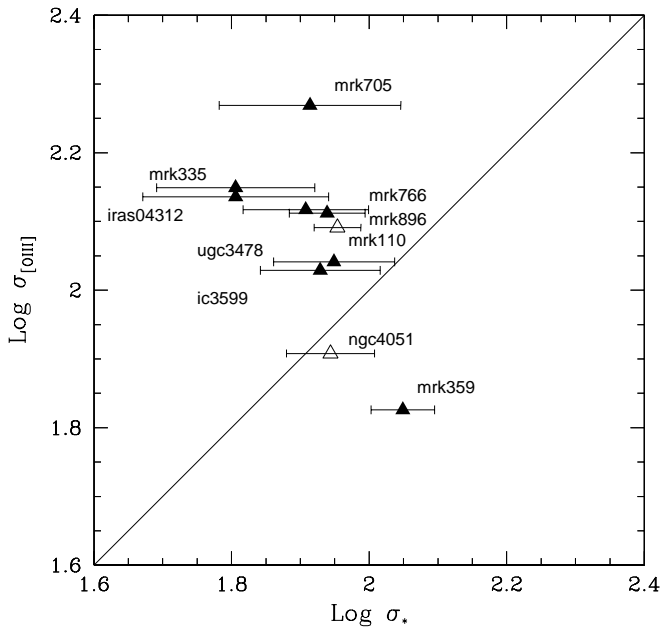, width=100mm}
 \caption{Stellar vs. gas kinematics. Solid triangles are the observed 
 galaxies, while open triangles the two additional NLS1s taken from 
 literature. The solid line is $\sigma_{[O~III]}=\sigma_*$.}
\label{fig2}
\end{center}
\end{figure}

In Fig.~\ref{fig2} we plotted $\sigma_*$ values vs. [O\,{\sc iii}] widths 
($\sigma_{\rm [O~III]}$).
The result is very interesting since it clearly shows that [O\,{\sc iii}]
sistematically overestimate the stellar velocity dispersion in NLS1s.
In fact all targets except one (Mrk 359) occupy the upper half of the 
$\sigma_*-\sigma_{[O~III]}$ plane, and are not grouped around the 1:1 
separation line. I
n other words, this is the first indication that using $\sigma_{\rm [O~III]}$
in place of $\sigma_*$ could be wrong in NLS1s, which seem to have a
$<\sigma_*> \sim 84\pm14$ \kms.
In this picture, Mrk~359 is an outlier. It is the only case among our data
having $\sigma_{[O~III]}<\sigma_*$. A justification was given by \citet{JB00}, who
suggested that such cases could be caused by gas settled in a cold rotating
disc.
 
\begin{figure}
\begin{center}
\psfig{file=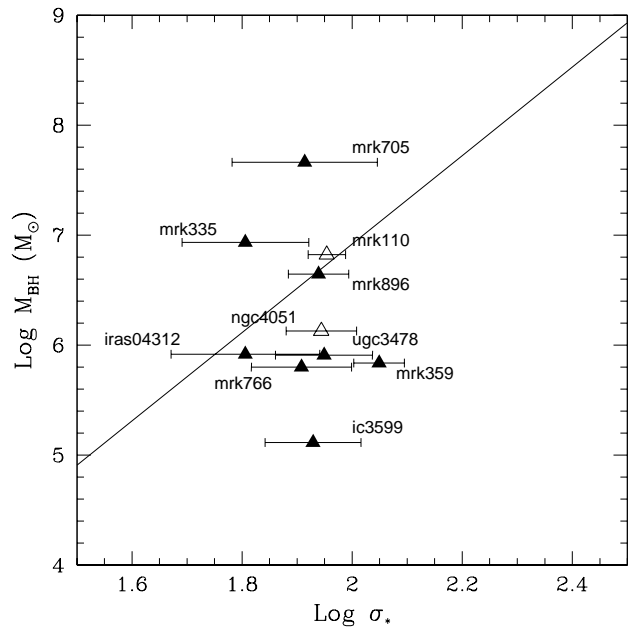, width=100mm}
 \caption{Stellar velocity dispersion vs. BH mass. The solid line is the 
 \citet{tret02} fit. Symbols are like in Fig.~\ref{fig2}.}
\label{fig3}
\end{center}
\end{figure}

Fig.~\ref{fig3} shows the new $\sigma_*$ data plotted vs. the BH masses. 
M$_{BH}$ values (Table \ref{tab2}) are taken from \citet{bo04} except for Mrk 766, 
which was observed also in the optical range between 4000 and 6000 \AA\ 
to measure the FWHM(H$\beta$) and the continuum luminosity at rest-frame
$\lambda=5100$ \AA\, as well as the [O\,{\sc iii}] width. 
Its $\sigma_{\rm [O~III]}$ and BH mass value are also listed in Table \ref{tab2}.
In both figures, open triangles indicate Mrk 110 and NGC 4051, two additional 
NLS1s with known $\sigma_*$, mentioned in Section 1.
Their $\sigma_{\rm [O~III]}$ and M$_{BH}$ values are taken from literature.
For Mrk 766 we used our measurements, which are however in agreement with 
\citet{JB00} within the error bars.

It is clear from Fig.~\ref{fig3} that these new $\sigma_*$ values fill a range 
more in agreement with the \citet{tret02} relation: 
$log~M_{BH}=4.02~log~\sigma_*-1.12$, considering a mass 
range of $\sim10^6-10^7~M_{\sun}$ for BHs in NLS1 galaxies.
In fact, after having plotted our data over the $M_{BH}-\sigma_*$ relation, we
see that NLS1s are well arranged around the Tremaine et al. fit, and not 
sistematically below the fit as previously found. 
The data show a significant scatter around the fit, not very different from
that observed in ordinary Seyfert 1 galaxies \citep[see e.g.][]{walu01, wan02,
bo04}. Moreover, 5 out of 10 NLS1s have a factor of 3-10 lower BH masses than 
predicted from the M$_{BH}-\sigma_*$ relation. This would suggest that NLS1s
have a tendency to lower M$_{BH}$/$\sigma_*$ ratios relative to
non-active galaxies and broad-line AGNs, but the limited number of points does 
not allow to draw a conclusion about this point.

\section{DISCUSSION AND CONCLUSIONS}

Several authors tested the M$_{BH}-\sigma_*$ correlation validity for AGNs
assuming that the gaseous component could reasonably trace the nuclear
gravitational potential. In particular, the [O\,{\sc iii}] width was used as 
surrogate of the stellar velocity dispersion, essentially for a practical 
reason: 
the [O\,{\sc iii}] emission line is usually bright in AGNs and therefore easily 
measurable even in case of distant objects. On the contrary stellar absorptions 
require longer integration time and/or large telescopes to reach the S/N ratio 
necessary to give a reliable estimate of the stellar velocity dispersion.
The FWHM([O\,{\sc iii}]) to $\sigma_*$ conversion was proposed by \citet{nel96},
who observed a large sample of AGNs in two spectral ranges, one including
H$\beta$, [O\,{\sc iii}] emission lines and Mg b triplet absorption, and the 
other containing the Ca\,{\sc ii} triplet absorption.
The data of 66 objects were plotted showing a moderate correlation with a
significant scatter.  
More recently \citet{JB00} obtained similar results on a
sample of Seyfert 1 and 2 galaxies, demonstrating that the majority of the 
galaxies form a cloud of points around the 1:1 line of the gas vs. stellar
velocity dispersion plot, with very weak or even no correlation. 

None the less, the use of the [O\,{\sc iii}] width in place of direct $\sigma_*$ 
measurements led \citet{bz04a}, \citet{gm04}, and \citet{bo04} to demonstrate that 
Seyfert 1 galaxies agree with the M$_{BH}-\sigma_*$ relation given by 
\citet{tret02}, even if with a large scatter of $\sigma_*$ (from 50 to more than 
500 \kms), while NLS1 galaxies do not. Moreover, NLS1s appear sistematically 
arranged below the Tremaine et al. fit. 
Interestingly, this result was already included in \citet{bo03}, who carried out 
an analysis of the $M_{BH}$--[O\,{\sc iii}] $\lambda$5007 width relation for 
107 low--redshift 
radio--quiet QSOs and Seyfert 1 galaxies and found that this correlation is 
real, but the scatter is large, about 0.2 dex.
Really, it is easy to verify that the narrow line AGNs (FWHM(H$\beta)<2000$ \kms) 
of his sample are mostly located below the Tremaine et al. fit.
Therefore, authors were induced to believe that NLS1s do not 
follow the BH-bulge correlation as Seyefert 1, quasars and normal galaxies.
But, very recently, \citet{bz04b} obtained a good consistency in BH mass 
estimates of NLS1 galaxies by using H$\beta$ and soft X-ray luminosity, while 
no agreement was found with the [O\,{\sc iii}] linewidth. 
Their result casts doubt on the 
reliability of [O\,{\sc iii}] as indicator of stellar velocity dispersion, at 
least for NLS1s. 
And, indeed it should be mentioned that the FWHM([O\,{\sc iii}])/2.35=$\sigma_*$ 
relation was obtained by \citet{nel96} for a sample of Seyfert 1 and 2 galaxies, 
but it was never verified for NLS1s.

The situation clearly required new direct $\sigma_*$ measurements.
We obtained these values for 8 NLS1s observing their Ca\,{\sc ii} triplet.
The choice of this spectral range was induced by the fact that NLS1s are very
often characterized by large Fe\,{\sc ii} emission multiplets (4400--4700 \AA\ 
and 5150--5350 \AA) in correspondence of the absorption lines commonly used to 
estimate $\sigma_*$, like Mg I $\lambda$5175 and Fe I $\lambda$5269. 
Unfortunately, as shown by \citet{pers88}, NLS1s with strong Fe\,{\sc ii} are 
likely to have Ca\,{\sc ii} in emission rather than in absorption.

By comparing stellar and gaseous kinematics we found that the 
FWHM([O\,{\sc iii}]) /2.35=$\sigma_*$ relation does not seem to hold for NLS1s, 
which are arranged mostly at values of $\sigma_{\rm [O~III]}$ larger 
than expected.
Contrary to the recent findings, the direct $\sigma_*$ measurements 
span a narrower range typically lower than 100 \kms, and such range is in 
agreement with the stellar velocity dispersions expected by the 
$M_{BH}-\sigma_*$ relation given by \citet{tret02} in case of 
$M_{BH}\sim10^6-10^7~M_{\sun}$. Indeed, with 
our $\sigma_*$ determinations, NLS1s are arranged around this relationship,
suggesting that in NLS1 galaxies
BH and bulge correlate, and lower mass BHs correspond to lower 
stellar velocity dispersions, or in other words lower mass bulges. 
We stress that this result is in agreement with our previous work \citep{bo04}, 
where we showed that NLS1s are mostly confined to the lower ranges of the 
M$_{BH}$-L$_{bulge}(B)$ plane, and are well correlated to Seyfert 1 galaxies and 
quasars. 
The fact that the stellar velocity dispersion in place of the gaseous one yields 
a tighter relation when correlated with BH mass, is essentially caused by the 
star kinematics which is more representative of the bulge
gravitational potential than the gas in AGNs. 
Even if the gas motion is mainly controlled by the mass of the bulge,
several observations in the past showed that the base of the line profiles
emitted by the Narrow Line Region (NLR) in many active galaxies is hardly 
reproduced simply assuming a pure virial motion in the potential 
of the host galaxy \citep[][ and references therein]{veil91}. 
Other important factors or processes can be involved, and these
additional processes can be at the origin of the larger scatter observed when
the FWHM([O III]) is used as surrogate of $\sigma_*$.
The interaction bewteen the gas and the ejected radio plasma observed in many
Seyfert galaxies may strongly influence the gas kinematics and contribute with
outflowing motions of the NLR clouds \citep{nel96}.
Moreover, supersonic winds generated by the active nucleus are believed to
accelerate the smaller clouds of the NLR located closer to the central engine 
and sometimes dominate their virial velocity \citep{smith93}.
This effect could be more important in NLS1 galaxies, since strong nuclear 
winds are expected in case of high accretion rates \citep{lh00}.
In addition, it was pointed out by \citet{nel96} that a slight tendency exists for 
barred and/or disturbed active galaxies to have broader 
[O\,{\sc iii}] emission lines. 
In conclusion, we obtained results with important implications, which strongly 
encourage additional observations to enlarge the sample of the observed 
galaxies.

\section*{Acknowledgements}
We are very grateful to the referee for precious comments
which improved the quality of the paper.
This research was partially based on data from the ING Archive.


\begin{thebibliography}{}

\bibitem[\protect\citeauthoryear{Bian \& Zhao}{2004a}]{bz04a} 
Bian W., Zhao Y., 2004a, MNRAS, 347, 607 

\bibitem[\protect\citeauthoryear{Bian \& Zhao}{2004b}]{bz04b}
Bian W., Zhao Y., 2004b, MNRAS, 352, 823

\bibitem[\protect\citeauthoryear{Boroson}{2003}]{bo03}
Boroson T.~A., 2003, ApJ, 585, 647

\bibitem[\protect\citeauthoryear{Botte et al.}{2004}]{bo04} 
Botte V., Ciroi S., Rafanelli P., Di Mille F., 2004, AJ, 127, 3168 

\bibitem[\protect\citeauthoryear{Di Matteo et al.}{2003}]{dima03} 
Di Matteo T., Croft R.~A.~C., Springel V., Hernquist L., 2003, ApJ, 593, 56

\bibitem[\protect\citeauthoryear{Ferrarese \& Merritt}{2000}]{fermer00} 
Ferrarese L., Merritt D., 2000, ApJ, 539, L9

\bibitem[\protect\citeauthoryear{Ferrarese et al.}{2001}]{fe01} 
Ferrarese L., Pogge R.~W., Peterson B.~M., 
Merritt D., Wandel A., Joseph C.~L., 2001, ApJ, 555, L79 

\bibitem[\protect\citeauthoryear{Gebhardt et al.}{2000a}]{gebet00a} 
Gebhardt K., et al., 2000, ApJ, 539, L13

\bibitem[\protect\citeauthoryear{Gebhardt et al.}{2000b}]{gebet00b} 
Gebhardt K., et al., 2000, ApJ, 543, L5 

\bibitem[\protect\citeauthoryear{Grupe \& Mathur}{2004}]{gm04} 
Grupe D., Mathur S., 2004, ApJ, 606, L41 

\bibitem[\protect\citeauthoryear{Jim{\' e}nez-Benito et al.}{2000}]{JB00} 
Jim{\' e}nez-Benito L., D{\'{\i}}az A.~I., Terlevich R., Terlevich E., 2000, 
MNRAS, 317, 907 

\bibitem[\protect\citeauthoryear{Kaspi et al.}{2000}]{ka00} 
Kaspi S., Smith P.~S., Netzer H., Maoz D., Jannuzi B.~T., Giveon U., 2000, 
ApJ, 533, 631 

\bibitem[\protect\citeauthoryear{Leighly \& Halpern}{2000}]{lh00} 
Leighly K., Halpern J., 2000, HEAD, 32, 1195 

\bibitem[\protect\citeauthoryear{Marconi \& Hunt}{2003}]{mahu03} 
Marconi A., Hunt L.~K., 2003, ApJ, 589, L21 

\bibitem[\protect\citeauthoryear{Mathur, Kuraszkiewicz, \& Czerny}{2001}]{matet01} 
Mathur S., Kuraszkiewicz J., Czerny B., 2001, NewA, 6, 321 

\bibitem[\protect\citeauthoryear{Nelson}{2000}]{ne00} 
Nelson C.~H., 2000, ApJ, 544, L91 

\bibitem[\protect\citeauthoryear{Nelson \& Whittle}{1995}]{nel95} 
Nelson C.~H., Whittle M., 1995, ApJS, 99, 67 

\bibitem[\protect\citeauthoryear{Nelson \& Whittle}{1996}]{nel96} 
Nelson C.~H., Whittle M., 1996, ApJ, 465, 96 

\bibitem[\protect\citeauthoryear{Ohyama et al.}{2002}]{Oh02} 
Ohyama Y., et al., 2002, AJ, 123, 2903 

\bibitem[\protect\citeauthoryear{Persson}{1988}]{pers88} 
Persson S.~E., 1988, ApJ, 330, 751

\bibitem[\protect\citeauthoryear{Shields et al.}{2003}]{sh03} 
Shields G.~A., Gebhardt K., Salviander S., 
Wills B.~J., Xie B., Brotherton M.~S., Yuan J., Dietrich M., 2003, ApJ, 
583, 124 

\bibitem[\protect\citeauthoryear{Smith}{1993}]{smith93} 
Smith S.~J., 1993, ApJ, 411, 570 

\bibitem[\protect\citeauthoryear{Tonry \& Davis}{1979}]{toda79} 
Tonry J., Davis M., 1979, AJ, 84, 1511 

\bibitem[\protect\citeauthoryear{Tremaine et al.}{2002}]{tret02} 
Tremaine S., et al., 2002, ApJ, 574, 740 

\bibitem[\protect\citeauthoryear{Veilleux}{1991}]{veil91} 
Veilleux S., 1991, ApJ, 369, 331 

\bibitem[\protect\citeauthoryear{Wandel}{2002}]{wan02} 
Wandel A., 2002, ApJ, 565, 762 

\bibitem[\protect\citeauthoryear{Wandel}{2004}]{wan04} 
Wandel A., 2004, Proc. of IAU Symp. 222, in press (astro-ph/0407399)

\bibitem[\protect\citeauthoryear{Wang \& Lu}{2001}]{walu01} 
Wang T., Lu Y., 2001, A\&A, 377, 52 

\bibitem[\protect\citeauthoryear{Whittle}{1992}]{whi92} 
Whittle M., 1992, ApJS, 79, 49 


\end{thebibliography}
\end{document}